# PbMnTeO$_6$: a chiral quasi 2D magnet with all cations in octahedral coordination and the space group problem of trigonal layered A$^{2+}$Mn$^{4+}$TeO$_6$

Mariia D. Kuchugura, [a,b] Alexander I. Kurbakov, [a,b] Elena A. Zvereva, [c] Tatyana M. Vasilchikova, [c] Grigory V. Raganyan, [c] Alexander N. Vasiliev,[c,d,e] Victor A. Barchuk[f] and Vladimir B. Nalbandyan *[f]

[a]NRC "Kurchatov Institute" – PNPI, 188300 Gatchina, Russia
[b]Faculty of Physics, St. Petersburg University, 199034 St. Petersburg, Russia
[c]Faculty of Physics, M.V. Lomonosov Moscow State University, Moscow 119991, Russia
[d]National Research South Ural State University, Chelyabinsk 454080, Russia
[e]National University of Science and Technology "MISiS", Moscow 119049, Russia
[f]Faculty of Chemistry, Southern Federal University, 344090 Rostov-on-Don, Russia. E-mail: vbn@sfedu.ru

Antiferromagnetic PbMnTeO$_6$, also known as mineral kuranakhite, has been reported recently to have all three cations in trigonal prismatic coordination, which is extremely unusual for both Mn(4+) and Te(6+). In this work, the phase was reproduced with the same lattice parameters and Néel temperature $T_N$ = 20 K. However, powder neutron diffraction unambiguously determined octahedral (trigonal antiprismatic) coordination for all cations within the chiral space group P312. The same symmetry was proposed for SrMnTeO$_6$ and PbGeTeO$_6$, instead of the reported space groups P-62m and P31m, respectively. PbMnTeO$_6$ was found to be a robust antiferromagnet with an assumingly substantial scale of exchange interactions since the Néel temperature did not show any changes in external magnetic fields up to 7 T. The determined effective magnetic moment $\mu_{eff}$ = 3.78$\mu_B$ was in excellent agreement with the numeri- cal estimation using the effective g-factor g = 1.95 directly measured here by electron spin resonance (ESR). Both specific heat and ESR data indicated the two-dimensional character of magnetism in the com- pound under study. The combination of chirality with magnetic order makes PbMnTeO$_6$ a promising material with possible multiferroic properties.

## 1. Introduction

Recently, the crystal structure and magnetic properties of PbMnTeO$_6$ have been reported, with all cations in trigonal-prismatic coordination within the space group P-62m. The non-centrosymmetric nature of the phase was demonstrated by the second harmonic generation (SHG) test.[1] Earlier, the structure of the same type was reported for SrMnTeO$_6$.[2] Although prismatic coordination is normal for large cations like Pb$^{2+}$ nd Sr$^{2+}$, it is extremely unusual for small cations like $Mn^{4+}$ and $Te^{6+}$, as evidenced from the database reviews reporting hundreds of oxygen octahedra, one exceptional $MnO_4$ tetrahedron and no prisms.[3,4] The O–O distances around Mn and Te are abnormally short, 2.135 and 2.144 Å, in $PbMnTeO_6$ and $SrMnTeO_6$, respectively.[1,2] Besides these geometrical constraints, a very high crystal field stabilization energy for $d^3$ cations in an octahedral environment is another argument against the prismatic coordination for $Mn^{4+}$. The adopted structural model for $AMnTeO_6$ (A = Pb and Sr) is especially strange compared with $AGeTeO_6$ (A = Sr[5] and Pb[6]), having essentially the same unit cell sizes (Table 1) but refined in the space group P312 with all cations in octahedral coordination.

The three space groups listed in Table 1 have an identical arrangement of cations and only differ in positions of oxygen and the order/disorder of M and Te. None of them generates systematic absences or superlattice reflection; hence, they may be distinguished only by a careful analysis of diffraction intensities. Since the location of light oxygen atoms by X-ray diffraction (XRD) in the presence of heavy atoms might be problematic, we decided to reinvestigate $PbMnTeO_6$ using neutron diffraction, in which the scattering power of oxygen is equal to that of Te and exceeds that of Mn. This seems important due to the following reasons.

(i) The space group P-62m is noncentrosymmetric but not chiral whereas P312 is chiral.[8] In combination with the antiferromagnetic ordering this may result in extremely interesting physical effects,[9,10] e.g., ferroelectric switching just by a slight tilting of the magnetic field direction.[10] For this reason, we also carried out detailed studies of the magnetic properties of $PbMnTeO_6$ by measuring the temperature dependences of magnetic susceptibility in magnetic fields, by a specific heat method and by the method of electron spin resonance.

(ii) Mn and Te in both the reported structures of $AMnTeO_6$ (A = Pb and Sr) are disordered over the same Wyckoff site. For neutron diffraction, the atomic scattering factors of Mn and Te are opposite in sign and this provides better opportunity for checking their order/disorder.

(iii) $PbMnTeO_6$ is also known as mineral kuranakhite.[11-14] Its low accuracy powder XRD patterns, except two or three weak reflections, are similar to those of the synthetic $PbMnTeO_6$. They were tentatively indexed to the C-centred orthorhombic cell with lattice parameters a = 5.11(1), b = 8.91 (2), c = 5.32(1) Å, and b/a = 1.74.[13] This strongly resembles an orthohexagonal setting of the cell reported in Table 1. Therefore, this work may also be important for mineralogy.

(iv) For $PbGeTeO_6$, instead of the non-polar space group P312,[6] a polar group was proposed recently[7] (Table 1). If it is correct, the same may be expected for $PbMnTeO_6$. Then, it would combine magnetic order with electrical polarity. Therefore, this point also requires elucidation.

## 2. Experimental
*2.1. Sample preparation and identification*
Polycrystalline $PbMnTeO_6$ was prepared by a slightly modified solid-state method.[1] Equimolar amounts of reagent-grade $PbO_2$, $MnO_2$ and $TeO_2$ were carefully mixed with a mortar and pestle, pressed into pellets and calcined in air at 450, 600 and 700 °C for 2 h each with intermediate regrinding and pressing. As in the preceding work,[1] the product was slightly contaminated with $Pb_2MnTeO_6$ which was easily removed by washing with diluted nitric acid followed by washing with water and drying.

The XRD pattern of the thus prepared black powder (Fig. 1) was recorded using an ARL X'TRA diffractometer. To avoid preferred orientation, the powder was mixed with equal weight (i.e., a much larger volume) of amorphous powder, instant coffee. Despite using a Si(Li) solid-state detector tuned for the measurement of CuK$\alpha$ radiation only, the strongest CuK$\beta$ reflection is still visible at 2$\Theta \approx 23.5°$ as shown in Fig. 1. Lattice parameters were refined by using the program CELREF 3 (J. Laugier and B. Bochu) and using corundum as an internal standard. They agree well with the previous data (Table 1). However, the agreement of relative intensities is not so good: the previously published pattern shows systematically enhanced 00$l$ and 10$l$ reflections, obviously due to grain orientation along the (001) layer plane.

Elemental analysis was done by the EDX method (see the ESI for details). It yielded the following gross formula: $Pb_{0.96(1)}Mn_{0.97(7)}Te_{1.00(3)}O_{4.9(5)}$. Within experimental uncertainties, the cationic composition is stoichiometric and agrees with the previously reported $Pb_{0.99(8)}Mn_{1.04(7)}Te_{0.98(1)}O_x$.[1]

Of course, oxygen determination by EDX is not reliable in the presence of heavy elements. To determine the y value in the formula $PbMnTeO_{5+y}$, i.e., oxygen excess over its content corresponding to Pb(2+), Mn(2+) and Te(6+), chemical redox analysis was employed.[15] The powdered samples were digested in aliquots of an aqueous $H_2SO_4$ + $FeSO_4$ solution. The resulting solutions were titrated with a standard $KMnO_4$ solution together with identical aliquots without tellurate. Then, the difference between the measured volumes of $KMnO_4$ was equivalent to y. This yielded the y value of 0.903(2), which is close to the ideal value of unity. We suppose that this slightly reduced result might be due to an admixture of Te(4+) and Mn (2+) in the form of tellurite glass, not detected by XRD but manifested in very strong sintering of the pellets. Another reason for the reduced result might be incomplete acid digestion due to the formation of insoluble $PbSO_4$ on the surface of $PbMnTeO_6$ particles.

*2.2. Neutron diffraction*

Neutron diffraction experiments were carried out at the spallation neutron source SINQ at PSI Villigen with the use of the instrument HRPT[16] at room temperature. Constant-wavelength neutrons ($\lambda$ = 1.494 Å) were produced by focusing a Ge monochromator with (335) reflection. The fixed take-off monochromator angle was 120°. The sample was placed into a cylindrical vanadium container with a diameter of 6 mm. The height of the sample was 50 mm. The instrumental parameters of the HRPT diffractometer were determined using the $Na_2Ca_3Al_2F_{14}$ standard. Structural refinements were performed by the Rietveld method using the FULLPROF suite.[17,18]

*2.3. Magnetic measurements*

The magnetic susceptibility measurements were performed using a Quantum Design SQUID

magnetometer MPMS-7T. The temperature dependence of the magnetic susceptibility was measured in the magnetic fields of B = 0.1 T and 7 T and in the temperature range of 1.8–300 K. The magnetization curve was recorded for the magnetic fields of B ≤ 7 T at T = 2 K after cooling the sample in zero field.

Electron spin resonance (ESR) studies were carried out using an X-band ESR spectrometer CMS 8400 (ADANI) (ν ≈ 9.4 GHz and B ≤ 0.7 T) equipped with a continuous He-gas flow cryostat, operating in the temperature range of 6 ≤ T ≤ 300 K. The effective g-factor has been calculated with respect to the BDPA (a,g-bisdiphenyline-b-phenylallyl) reference sample with $g_{et}$ = 2.00359.

*2.4. Specific heat*

Specific heat studies were carried out by a relaxation method using a Quantum Design PPMS-9T system on the cold-pressed $PbMnTeO_6$ and non-magnetic analogue $PbGeTeO_6$ samples. The data were collected at zero magnetic field in the temperature range of 2–300 K.

## 3. Results and discussion

*3.1. Crystal structure*

Fig. 2a represents the results of neutron diffraction refinement based on the published model with the space group P-62m.[1] It is evident that the calculated profile strongly disagrees with the experimental points. Then, an alternative model was considered based on the $PbSb_2O_6$ (rosiaite) type. It has the same arrangement of cations as in the published structure (including the Mn/Te disorder), but with different oxygen positions, resulting in octahedral (trigonal antiprismatic) coordination of all cations. This model, P-31m, looks much better (Fig. 2b and Table 2) but it is centrosymmetric and, thus, contradicts the SHG test.[1] Finally, the Mn/Te ordering in octahedral sites was introduced by analogy with $AGeTeO_6$ (A = Sr[5] and Pb[6]), the space group P312, and this model resulted in the best fit (Fig. 2c and Table 2).

At this point we noticed most recent publication[7] where crystal structure of $PbGeTeO_6$ was refined in polar space group P31m.

This model differs from that reported earlier[6] in terms of three features: 0.7–0.8% bigger lattice parameters (Table 1), Ge/Te disorder, and weak polar displacements of all cations relative to anions (or vice versa) by only 0.027 Å along the three-fold axis. All of this seems strange. The large difference in lattice parameters might be due to the experimental errors but might also reflect the real difference in the structure and/ or composition because the solid-state preparation of non- polar $PbGeTeO_6$ samples was made at 700 °C (ref. 6) or 730 °C (in this work, see the ESI) whereas the P31m crystals were reportedly grown using the $TeO_2$ flux at 600 °C and below.[7] It is, however, unnatural that lower preparation temperature in the presence of flux results in cation disorder[7] whereas the high-temperature product is cation-ordered.[6] The polar displacement of $Pb^{2+}$, in contrast to $Sr^{2+}$, might be due to the stereochemically active lone electron pair of $Pb^{2+}$ but the same displacements of $Mn^{4+}$ and $Te^{6+}$ seem strange. Moreover, such small static displacements cannot be proved definitely by diffraction methods alone due to strong correlation with thermal displacement parameters. If, nevertheless, such small displacements are real, they must be easily reoriented by an external electric field, i.e., PbGeTeO$_6$ should be ferroelectric, and the same may be expected for PbMnTeO$_6$.

To check these possibilities, we performed dielectric studies of PbGeTeO$_6$ (see Section 3.2) and neutron diffraction refinement of the polar model for PbMnTeO$_6$ (Fig. 2d). Here, the space group P31m[7] could not be applied since an ordered Mn/Te arrangement was proved earlier in the P312 model; then, the suggested polarity required the elimination of the two-fold axis leading to the space group P3. Refinement within this model, despite the much larger number of variables, did not bring considerable improvement with respect to the P312 model (Table 2). Therefore, we consider the P312 model as the most correct description of PbMnTeO$_6$. Note that even the dielectric studies of the Ge analogue (Section 3.2) could not confirm its polarity.

Refinement details, atomic coordinates, site occupancies and displacement parameters for PbMnTeO$_6$ are listed in Tables S2 and S3 of the ESI, and the important interatomic distances are given in Table 3. The refined crystal structure is compared with the previously reported model in Fig. 3. The corresponding crystallographic information file (CIF) has been deposited with the CCDC, database code 1867240.

The refinement resulted in 9.4(6) % Mn/Te inversion. However, we suppose that this may be a fictitious effect due to the faulted stacking of completely ordered individual layers. The bond distances agree well with the corresponding sums of ionic radii (Table 3). It should be emphasized that abnormally short O–O distances are avoided in the octahedral structure, the shortest O–O distance being 2.46 Å (Table 3). The crystal structure is based on the cation-ordered (MnTeO$_6$)$^{2-}$ layers of edge-shared MnO$_6$ and TeO$_6$ octahedra, with Pb$^{2+}$ cations between these layers, also in oxygen octahedra. The previous model[1] differs in terms of two main features: Mn/Te disorder and prisms in place of all octahedra. Interestingly, the cationic environment of oxygen is essentially the same in both the old and new structural models: almost planar with the sum of three bond angles of 353.5 and 354.2°, respectively.

It is now evident that SrMnTeO$_6$ should also have an octahedral, rather than prismatic, structure. Still, there are two possibilities: it may be either strictly isostructural with PbMnTeO$_6$ and SrGeTeO$_6$, i.e. chiral with an ordered Mn and Te, or the centrosymmetric rosiaite-type with disordered Mn and Te.

*3.2. Preparation and dielectric studies of PbGeTeO$_6$*

PbGeTeO$_6$ was prepared for two purposes. Its powder was used as a non-magnetic analogue for specific heat measurements (see Section 3.3) and its ceramic form for dielectric measurements. Details of its preparation and characterization are described in the ESI. Its phase purity is illustrated by the XRD pattern (Fig. S1 of the ESI), its refined lattice parameters are in excellent agreement with the initial report[6] and differ significantly from the most recent single-crystal data[7] (Table 1). Uniaxial hot pressing yielded 95% dense ceramic discs with preferred orientation of the three-fold axis normal to the disc (Fig. S2), which is favourable for the discovery of dielectric anomalies, if any. However, no signs of the expected ferroelectric phase transition could be found between room temperature and 300 °C (Fig. S3), and the values of relative dielectric permittivity, e.g., 21 at room temperature, were too low for dense ferroelectric ceramics. Very small polar displacements reported for the P31m model[7] predict high polarizability and low Curie point, which is contrary to our observations. Moreover, although Jia et al. reported the crystal growth of $PbGeTeO_6$ "from a spontaneous method using $TeO_2$ as flux" starting from 600 °C,[7] the mixture of $PbGeTeO_6$ with $TeO_2$ in our experiments remained completely solid even at 630 °C. After a 3 h annealing at this temperature followed by slow cooling, the XRD pattern showed a mixture of $TeO_2$ with the trigonal phase having essentially the same lattice parameters as those given in Table 1: a = 5.0875(2) and c = 5.4478(1) Å.

Therefore, we must conclude that the report on $PbGeTeO_6$ P31m single crystals with the expanded unit cell[7] is incorrect and the compound is non-polar, in accordance with the initial report.[6]

*3.3. Static and dynamic magnetic properties*

The magnetic susceptibility data $\chi$ = M/B of $PbMnTeO_6$ are presented in Fig. 4. No difference was observed between the zero- field-cooled (ZFC) and field-cooled (FC) susceptibilities, signalling the absence of either spin glass or cluster glass effects. One can see a clear kink on $\chi(T)$ at about 20 K on both ZFC and FC curves. Obviously, it can be assigned to the onset of long-range antiferromagnetic (AFM) order in agreement with our specific heat data (see below) and the previously reported data.[1] In the paramagnetic state, $\chi(T)$ can be approximated by the Curie–Weiss law with the inclusion of the temperature- independent term $\chi_0$, i.e. $\chi = \chi_0 + C/(T - \Theta)$, where C is the Curie constant and $\Theta$ is the Weiss temperature. We should note that the authors of the earlier work[1] did not take into account the temperature-independent term during the analysis of the magnetic susceptibility, while it strongly affects the correct determination of the Weiss temperature, especially for frustrated and low-dimensional (LD) systems in which short- range correlations exist over a wide temperature range above the ordering transition. This leads to a strong deviation from the Curie–Weiss law at temperatures essentially higher than $T_N$ and requires an accurate procedure for choosing the correct range for quantitative analysis. Indeed, as one can see from Fig. 4, the experimental data essentially deviate down- ward from the extrapolation of the fitting curve at lower temperatures. Such behaviour evidences the strengthening of the contribution of antiferromagnetic interactions. It is most clearly seen from the C versus T dependence shown in the lower panel of Fig. 4. This representation of the experimental data is most suitable for LD magnetic systems in which the presence of the temperature-independent term can significantly distort the temperature dependence of the inverse mag- netic susceptibility and reduce the reliability of analysis.

Hence, the FC curve was fitted in the temperature range of 200–300 K. The parameters yielded from the approximation are $\chi_0$ = 4.37×10$^{-5}$ emu mol$^{-1}$, C = 1.79 emu K mol$^{-1}$ and $\Theta$ = −38 ± 1 K.

The positive value of $\chi_0$ indicates the predominance of the van Vleck contribution over the sum of the Pascal constants of individual ions in $PbMnTeO_6$ $\chi_{dia}$ = −1.24 × 10$^{-4}$ emu mol$^{-1}$.[20]

Such van Vleck contribution can arise from the splitting of the d-shell of $Mn^{4+}$ ions in a distorted octahedral environment (although Mn–O distances are identical, O–Mn–O bond angles vary between 80.1 and 97.6°). The negative sign of Θ indicates the prevailing antiferromagnetic exchange interactions at elevated temperatures. Despite the triangular arrangement of the magnetic subsystem (see the upper inset in Fig. 5), the frustration is relatively low (f = |Θ|/$T_N$ ~ 2) probably due to the fact that exchange interactions between the Mn ions occur via long super-exchange paths Mn–O–Te–O–Mn. We can also conclude that the scale of exchange interactions should be substantial since the value of the Néel temperature does not change in the applied external magnetic field of B = 7 T (the lower inset in Fig. 5).

The value of C determined in the range of 200–300 K gives the effective magnetic moment of $Mn^{4+}$ ions $\mu_{eff}$ = 3.78$\mu_B$. In order to determine the effective g-factor aiming to estimate the theoretical value of the effective magnetic moment in accordance with $\mu_{theor}$ = $[g^2\mu_B^2 S(S+1)]^{1/2}$, we carried out the electron spin resonance (ESR) measurements of $PbMnTeO_6$ (see the inset in the upper panel of Fig. 4). The fitting of the ESR absorption line gives the value of the effective g-factor as g = 1.95 ± 0.01. This value is markedly lower than the spin-only value of 2 and implies a non-negligible contribution of the orbital moment as indeed one can expect for the $Mn^{4+}$ ions with less than half filling of the electron d-shell ($d^3$). As a consequence, the corresponding theoretical value of the magnetic moment $\mu_{theor}$ = 3.78$\mu_B$ obtained here is in excellent agreement with the experimental one in contrast to the previous data.[1]

The M(B) isotherm at T = 2 K for $PbMnTeO_6$ shows neither hysteresis nor saturation in the magnetic fields up to 7 T (the main panel of Fig. 5). There is a slight bending in the M vs. B curve, which is weaker than expected for an S = 3/2 magnet. Such a type of magnetization curve is characteristic of LD magnets with strong short-range AFM correlations and frustration. It is also noted that within this range of the applied magnetic fields, the magnetic moment is still noticeably lower than the theoretically expected saturation magnetic moment for the $Mn^{4+}$ ion, $M_S$ = $gS\mu_B$ = 2.9$\mu_B$.

Specific heat data in zero magnetic field are in good agreement with the temperature dependence of the magnetic susceptibility in weak magnetic fields, demonstrating the distinct λ-type anomaly, confirming the onset of the magnetic order at $T_N$ = 20 K (Fig. 6). The position of this anomaly on $C_p(T)$ perfectly coincides with the maximum on the Fisher specific heat[21,22] d(χT)/dT(T) that is characteristic of LD antiferromagnets with strong short-range correlations (the upper inset in Fig. 6). For quantitative estimations, the specific heat data were also measured for the nonmagnetic isostructural analogue $PbGeTePO_6$. One can assume that the specific heat of this isostructural compound provides an estimate for the pure lattice contribution to specific heat of the studied compound. The correction to this contribution for $PbMnTeO_6$ has been made taking into account the difference between the molar masses of Ge and Mn atoms.[23] The values for the Debye temperature $\Theta_D$ have been estimated to be 247 ± 5 K for the diamagnetic compound $PbGeTeO_6$ and 250 ± 5 K for the compound $PbMnTeO_6$.

It is established that the specific heat jump at Néel temperature amounts to $\Delta C_m$ ≈ 10.5 J (mol K)$^{-1}$ for $PbMnTeO_6$ (the lower inset in Fig. 6), which is lower than the values predicted from the mean-field theory for the antiferromagnetic spin ordering assuming manganese to be in the high-spin state $Mn^{4+}$ (S = 3/2) ($\Delta C_m = 5R \cdot S(S+1)/[S^2 + (S+1)^2] \approx 18.3$ J (mol K)$^{-1}$).[23] This indicates the presence of appreciable short-range correlations far above $T_N$, which is typical for frustrated and LD systems.[24] We analyzed the magnetic counterpart to the specific heat $C_m(T)$ below $T_N$ using the spin-wave (SW) approach assuming that the limiting low-temperature behaviour of the magnetic specific heat should follow the $C_m \sim T^{d/n}$ power law for magnons,[25] where d stands for the dimensionality of the magnetic lattice and n is defined as the exponent in the dispersion relation $\omega \approx \kappa^n$. For PbMnTeO$_6$, the least-squares fitting of the data below $T_N$ (the red solid line in the lower inset of Fig. 6) achieved good accuracy, d = 2± 0.1 and n = 1 ± 0.1 values, which implies the presence of 2D AFM magnons at the lowest temperatures. It was found that the magnetic entropy $\Delta S_m$ saturates at about 20 K reaching approximately $\Delta S_m \approx 8$ J (mol K)$^{-1}$ which is significantly lower than those expected from the mean-field theory for the S = 3/2 spin system ($\Delta S_m = 2R \ln(2S + 1) \approx 15.6$ J (mol K)$^{-1}$).[23] This reveals the existence of noticeable frustration and short-range correlations far above $T_N$, which is a characteristic feature of 2D magnetism for PbMnTeO$_6$.[24]

Over the whole temperature range, the ESR spectra reveal a single resonance line, which can be perfectly described by the standard Lorentzian profile (the inset in Fig. 4). The integral ESR intensity $\chi_{ESR}$ was obtained by double integration of the first derivative absorption line and is shown in the upper panel of Fig. 4 along with the static magnetic susceptibility for comparison. The effective g-factor and the ESR linewidth $\Delta B$ derived from the Lorentzian fits for the ESR spectra are shown in Fig. 7.

The effective g-factor remains almost temperature-independent down to 175 K followed by a slight deviation from the value g ~ 1.95 with decreasing temperature. The ESR linewidth shows a continuous increase as the temperature decreases. Such an increase in $\Delta B$ is usually observed in antiferromagnets due to the slowing down of spin fluctuations as the critical temperature is approached from above.[26-28] The temperature variation of $\Delta B$ can be described as $\Delta B(T) = \Delta B^* + A(T_N^{ESR}/(T - T_N^{ESR}))^\beta$, where A denotes an empirical parameter, $\Delta B^*$ is the high-temperature limiting the minimum value of the linewidth, $T_N^{ESR}$ is the temperature of the order–disorder transition, and β is a critical exponent. A least squares fitting of the experimental data for PbMnTeO$_6$, shown by the solid red line in Fig. 7, gives the following parameters: $T_N^{ESR}$ = 19.9 ± 1 K and β = 1.24 ± 0.1. Evidently, the value of $T_N^{ESR}$ is close to the Néel temperature $T_N$ derived from the static magnetic measurements. The observed critical exponent is close to the values for 2D oxides.[29]

## 4. Conclusions

The previously reported P-62m structure of PbMnTeO$_6$ with cations in prismatic coordination and abnormally short O–O distances has been revised using powder neutron diffraction. It is shown unambiguously that PbMnTeO$_6$ belongs to the chiral space group P312 with all cations in octahedral coordination and an almost planar cationic environment of oxygen. Highly likely, this model should also be valid for SrMnTeO$_6$ and PbGeTeO$_6$; a recent report on the polar structure of PbGeTeO$_6$ is dubious and not supported by dielectric measurements. Both magnetic susceptibility and specific heat have detected antiferromagnetic ordering in

PbMnTeO$_6$ below the Néel temperature of 20 K and this value is in agreement with previously reported data. We have also found that appropriate analysis of magnetic susceptibility requires the inclusion of the temperature-independent contribution. ESR spectroscopy gives the value of the effective g-factor as g = 1.95, which indicates the non-negligible contribution of the orbital moment for Mn$^{4+}$ (d$^3$) ions. As a consequence, the corresponding value of the magnetic moment obtained from the Curie constant here agrees much better with the theoretical one in contrast to the previous result.[1] Our data in applied magnetic fields imply a large scale of magnetic super-exchange interactions and moderate frustration. Besides, we have shown an essential role of short-range correlations triggered by the quasi-two dimensional layered structure and the triangular arrangement of the magnetic sublattice of Mn$^{4+}$ ions. The discovery of chirality in the antiferromagnetic PbMnTeO$_6$ makes it a potentially promising material with unusual magnetic and perhaps multiferroic properties.

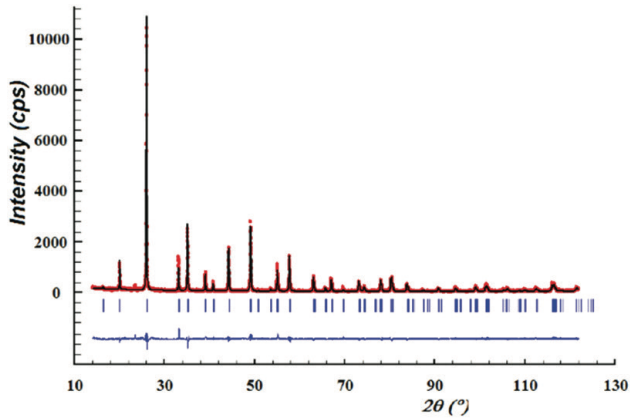

**Fig. 1.** Experimental (red points) and theoretical (black line) XRD pat- terns of PbMnTeO$_6$. The theoretical pattern was calculated using the structural data from the neutron refinement (see Section 3.1), and only instrumental and profile parameters were refined to fit the XRD data. The blue line in the bottom is the difference profile.

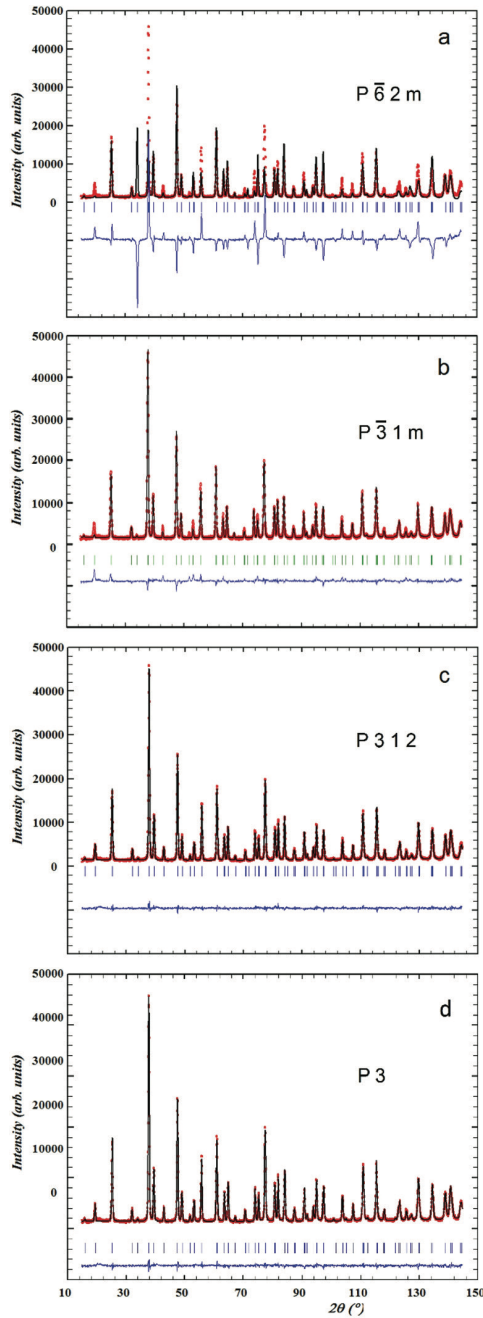

**Fig. 2.** Four variants of the Rietveld refinement of the neutron powder diffraction pattern of PbMnTeO$_6$. The calculated (continuous lines) and experimental (circles) intensities are plotted. Tick marks correspond to the Bragg positions. The bottom lines are the difference curves.

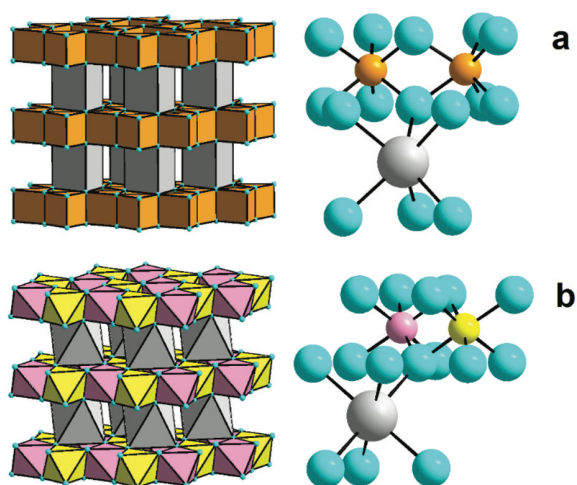

**Fig. 3.** Comparison of the two structural models for PbMnTeO$_6$: previously published (a) and revised (b) in polyhedral (left) and ball-and-stick (right) presentation. Cyan balls, oxygen; grey balls and polyhedra, Pb; orange balls and prisms, (Mn$_{0.5}$Te$_{0.5}$); pink balls and octahedra, (Mn$_{0.91}$Te$_{0.09}$); and yellow balls and octahedra, (Te$_{0.91}$Mn$_{0.09}$).

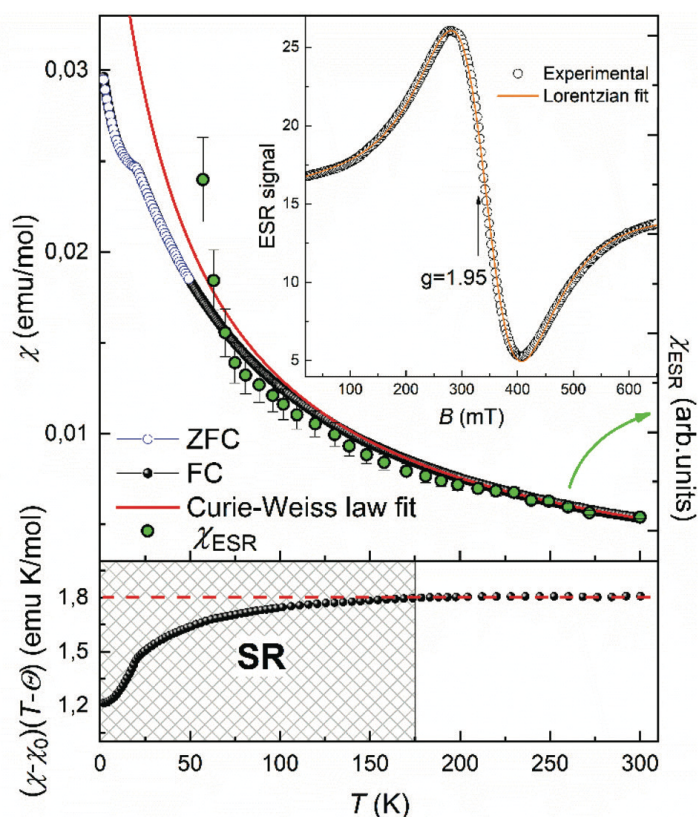

**Fig. 4.** The temperature dependence of the magnetic susceptibility of PbMnTeO$_6$ in both FC and ZFC regimes (upper panel) along with the dynamic magnetic susceptibility $\chi_{ESR}$ obtained from the ESR data (green filled symbols), the inset in this panel represents the ESR spectrum of the title compound taken at room temperature. The temperature dependence of the Curie constant $C = (\chi - \chi_0)(T - \Theta)$ is shown in the lower panel. The horizontal dashed line in the lower panel represents the limiting value $C = 1.79$ emu K mol$^{-1}$.

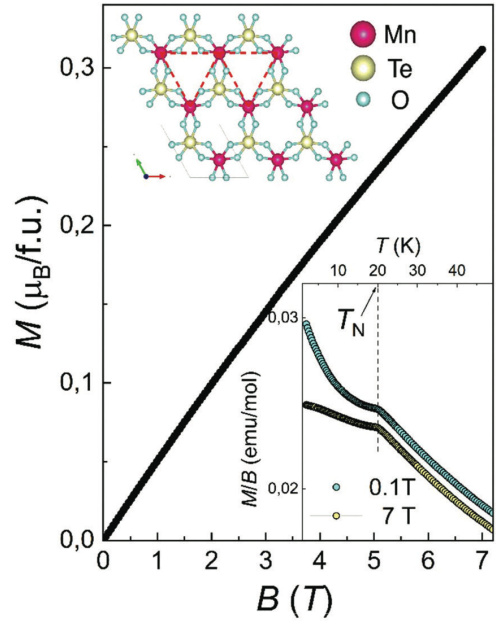

**Fig. 5.** The magnetization isotherm of PbMnTeO$_6$ at T = 2 K. Insets: a triangular motif of Mn$^{4+}$ ions in magnetically active layers (upper); the temperature dependencies of reduced magnetization at B = 0.1 and 7 T (lower).

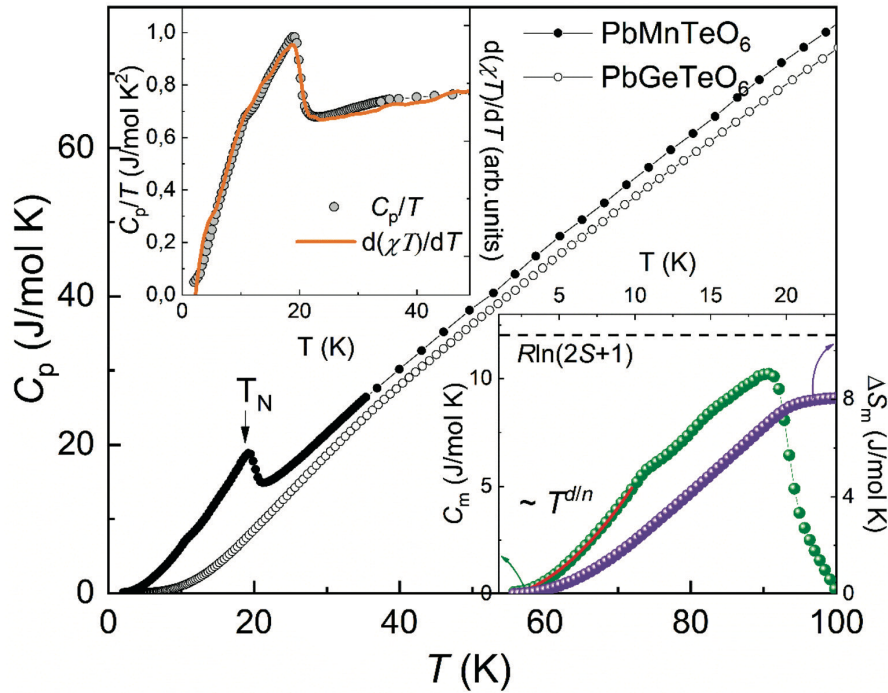

**Fig. 6.** The temperature dependence of the specific heat of PbMnTeO$_6$ (black filled symbols) and the non-magnetic isostructural analogue PbGeTeO$_6$ (black open symbols) in zero magnetic field. Upper inset: C/T(T) dependence (grey symbols) along with Fisher specific heat d($\chi$T)/dT(T) (orange line). Lower inset: the temperature dependences of C$_m$(T) (green symbols) and the entropy S$_m$(T) (violet symbols). The red line is the result of approximation in the framework of the spin wave theory.

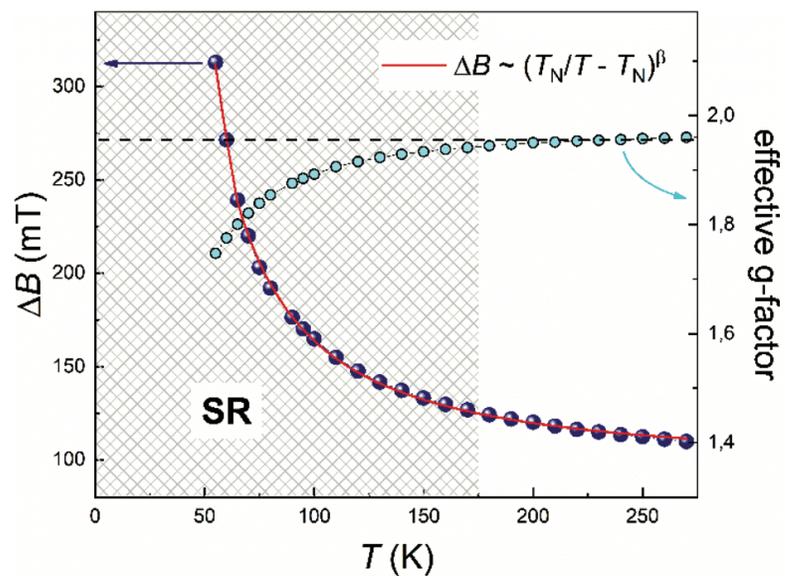

**Fig. 7.** The temperature dependencies of the effective g-factor and the ESR linewidth ΔB for PbMnTeO$_6$ derived from the ESR data. The solid line represents approximations in accordance with the power law (see text).

**Table 1.** Comparison of crystallographic data for AMTeO$_6$ (A = Pb and Sr; and M = Mn and Ge) from different X-ray diffraction (XRD) studies

| Composition | Space group | a, Å | c, Å | V, Å$^3$ |
|---|---|---|---|---|
| SrMnTeO$_6$ [2] | P-62m | 5.1425(9) | 5.384(2) | 123.3 |
| PbMnTeO$_6$ [1] | P-62m | 5.10143(5) | 5.39643(6) | 121.6 |
| Same, this work | P312 | 5.1010(2) | 5.3951(2) | 121.6 |
| SrGeTeO$_6$ [5] | P312 | 5.06566(3) | 5.40394(5) | 120.1 |
| PbGeTeO$_6$ [6] | P312 | 5.08939(1) | 5.44883(4) | 122.2 |
| PbGeTeO$_6$ [7] | P31m | 5.131(8) | 5.486(11) | 125.1 |
| Same, this work |  | 5.0885(2) | 5.4487(1) | 122.2 |

**Table 2.** Comparison of discrepancy factors for the four structural models of PbMnTeO$_6$

| Space group | P-62m | P-31m | P312 | P3 |
|---|---|---|---|---|
| Figure | 2a | 2b | 2c | 2d |
| $\chi^2$ | 38.2 | 4.78 | 1.46 | 1.47 |
| $R_p$ | 19.5 | 6.23 | 4.17 | 4.18 |
| $R_{wp}$ | 27.1 | 9.89 | 5.47 | 5.5 |
| $R_{exp}$ | 4.38 | 4.52 | 4.53 | 4.54 |

**Table 3.** Interatomic distances in PbMnTeO$_6$, space group P312, compared with sums of ionic radii[19]

|  | L, Å | Sum of radii, Å |
|---|---|---|
| Pb–O | 2.5339(17) × 6 | 2.55 |
| (Mn$_{0.91}$Te$_{0.09}$) – O | 1.9126(12) × 6 | 1.895 |
| (Te$_{0.91}$Mn$_{0.09}$) – O | 1.9255(18) × 6 | 1.915 |
| O–O (shortest only) | 2.4613(15) × 1, 2.7469(12) × 2 |  |